# A Self-Correcting Strategy of the Digital Volume Correlation Displacement Field Based on Image Matching: Application to Poor Speckles Quality and Complex-Large Deformation

C.S. Li[a,b,*]; Z.J. Liu[a]

[a] (Department of Civil Engineering and Smart Cities, Shantou University, Shantou, 515000, China)

[b] (State Key Laboratory of Geomechanics and Geotechnical Engineering, Institute of Rock and Soil Mechanics, Chinese Academy of Sciences, Wuhan, Hubei 430000, China)

*Corresponding author.

E-mail address:

chengshengli@stu.edu.cn (C.S. Li)

22zjliu@stu.edu.cn (Z.J. Liu)

**Keywords**: Digital volume correlation; Displacement field correction; Poor speckle quality; Complex-large deformation; Micro-CT

**Abstract:**

**Background** Digital Volume Correlation (DVC) is widely used for the analysis of three-dimensional displacement and strain fields based on CT scans. However, the applicability of DVC methods is limited when it comes to geomaterials: CT speckles are directly correlated with the material's microstructure, and the speckle structure cannot be artificially altered, with generally poor speckle quality. Additionally, most geomaterials exhibit elastoplastic properties and will undergo complex-large deformations under external loading, sometimes leading to strain localization phenomena. These factors contribute to inaccuracies in the displacement field obtained through DVC, and at present, there is a shortage of correction methods and accuracy assessment techniques for the displacement field.

**Objectives** If the accuracy of the DVC displacement field is sufficiently high, the gray residue of the two volume images before and after deformation should be minimal, utilizing this characteristic to develop a correction method for the displacement field is feasible.

**Methods** The proposed self-correcting strategy of the DVC displacement field based on image matching, which from the experimental measurement error.

**Results** We demonstrated the effectiveness of the proposed method by CT triaxial tests of granite residual soil. Without adding other parameters or adjusting the original parameters of DVC, the gray residue showed that the proposed method can effectively improve the accuracy of the displacement field. Additionally, the accuracy evaluation method can reasonably estimate the accuracy of the displacement field.

**Conclusion** The proposed method can effectively improve the accuracy of DVC three-dimensional displacement field for the state of speckles with poor quality and complex-large deformation.

# Introduction

As multiscale material mechanics continues to evolve, current research is shifting its focus from macroscopic mechanics to the microscopic mechanisms of materials. This transition has created a growing need for corresponding microscopic experimental measurements, particularly in the field of non-destructive measurement of internal material structures using micro/nano X-CT technology and associated displacement field measurement techniques. In 1999, Bay et al. [1] extended the Digital Image Correlation (DIC) method to three dimensions, introducing a method known as Digital Volume Correlation (DVC) for calculating three-dimensional displacement fields based on CT scanning. Over the years, DVC has undergone rapid development and has found applications in various fields [2], such as biomaterial mechanics [3], concrete [4], composite materials [5], and nanoscale indentation [6]. Moreover, in geotechnical engineering, DVC has proven to be valuable not only for studying the failure mechanisms of rocks and patterns of strain localization [7, 8] but also for analyzing crack evolution and other soil mass phenomena [9, 10].

However, when applying Digital Volume Correlation (DVC), it is important to be aware of two potential issues that can impact the accuracy of the displacement field. The first issue involves the quality of the CT speckle pattern of the material. CT volume images are generated after the reconstruction of the CT scanning process, and the distribution of CT speckles' grayscale is directly linked to the material's microstructure. This differs significantly from artificially created DIC two-dimensional speckles or surface speckles, as CT speckles are inherent to the material and cannot be modified artificially. The second issue arises when elastoplastic materials undergo external loading, resulting in complex and substantial deformation, including strain localization such as shear bands. Sometimes, the strain within a shear band may exceed 0.3 to 0.5, leading to decorrelation problems. If both of these issues occur simultaneously during DVC calculation, they can present significant challenges to the accuracy of the displacement field obtained.

Many scholars have conducted extensive research on the accuracy and precision of DVC. In terms of speckle quality research, Park et al. [11] used the standard deviation of gray intensities within each speckle is introduced as a new metric to assess the quality of speckle patterns, proposed a speckle-pattern quality measurement, which integrates the features of gray intensity and speckle morphology. Su et al. [12] systematically analyzed the performance of digital speckles in terms of uniqueness, accuracy, precision, and spatial resolution, and proposed a theoretical model for analyzing systematic errors and random errors, which can ultimately enhance the measurement capabilities of the digital image correlation technique. Shao et al. [13] optimized the selection of areas with better speckle quality for DIC calculation to improve computational efficiency and accuracy. Wang et al. [14] studied the optimal aperture and digital speckle optimization methods in digital image correlation from an experimental measurement perspective. Su [15] established a theoretical model characterizing the low-pass filtering effect, which can consider the subset size for measuring displacement, the order of shape functions, and the influence of the underlying deformation field. The study focused on the impact of low-pass filtering effects in digital image correlation caused by shape function mismatch. Moreover, Croom et al. [16] and Lauren et al. [17] studied the impact of CT imaging on the accuracy of DVC.

In the study of complex strain fields. For global DVC method, Wittevrongel et al. [18] principled from p-adaptive finite element analysis were implemented to obtain a self-adapting higher order mesh, to accurately determine small strains with high strain gradients. Yang et al. [19] proposed augmented Lagrangian digital volume correlation (ALDVC), which combines the advantages of both the local (fast computation time) and global (compatible displacement field) methods, which has high accuracy and precision while maintaining low computational cost. And Yang et al. [20] proposed a spatiotemporally adaptive quadtree mesh (STAQ) DIC method, which could has high accuracy in solving complex geometric and discontinuous deformation fields in an automated fashion. Rouwane et al. [21-23] proposed an architecture-driven DVC metho based on automatic finite-element by a weak elastic regularization, which could calculated the field of material without or lack of CT speckle, effectively measure the three-dimensional strain field of honeycomb structural materials. For local DVC method, Lu and Cary [24] and Lan et al. [25] used a second-order form function can effectively improve the matching degree of subvolume. Wang and Pan [26, 27] proposed a self-adaptive DVC method, which could self-adaptive selection of optimal subvolume size and the best shape function is therefore highly desirable to realize full-automatic and quality DVC measurements. And Pan and Zou [28] also proposed a quasi-gauss point DIC/DVC methos, which could reduce systematic errors due to undermatched shape functions. Moreover, Ghulam Mubashar Hassan [29] used pattern matching method to measure the calculation of non-continuous displacement fields.

Research indicates that high-quality speckle patterns exhibit characteristics such as isotropy, non-periodicity, rich details, and good contrast. However, achieving satisfactory speckle quality in CT volume images of many materials remains challenging. This is particularly problematic when analyzing three-dimensional strain fields in plastic materials, where the quality of CT speckles is often subpar and deformations are complex and extensive. Current DVC methods struggle to provide accurate displacement fields and lack targeted displacement field correction techniques. Furthermore, existing experimental measurement methods (e.g., laser, strain gauges, optical fibers) make it difficult to non-destructively obtain true three-dimensional displacement or strain fields within samples. This limitation hinders the ability to correct and validate the precision of DVC displacement fields using these auxiliary methods. Therefore, in response to the demand for displacement field analysis applications with poor speckle quality and complex and large deformations, we have established a self-repair strategy for DVC displacement fields based on image matching properties of volume images before and after deformation, to provide methods and evaluation recommendations for improving the accuracy of DVC displacement fields.

# Self-Correcting Strategy of DVC Displacement Field

## Correction method for displacement field

The advancement of material research has led many researchers to utilize non-destructive CT scanning technology for studying the microstructural deformation or failure mechanisms of materials experiencing complex and extensive deformations. For instance, by combining CT scanning with load tests, researchers can explore the localized strain evolution mechanisms of materials from a microscopic perspective. However, as illustrated in Fig. 1, the CT speckles observed in the material are inherently linked to its material composition, differing significantly from artificially generated 2D speckles. The quality of CT volume image speckles cannot be artificially manipulated, making it challenging to enhance their quality. When materials undergo complex and substantial deformations under external loading, traditional DVC methods may result in decorrelation.

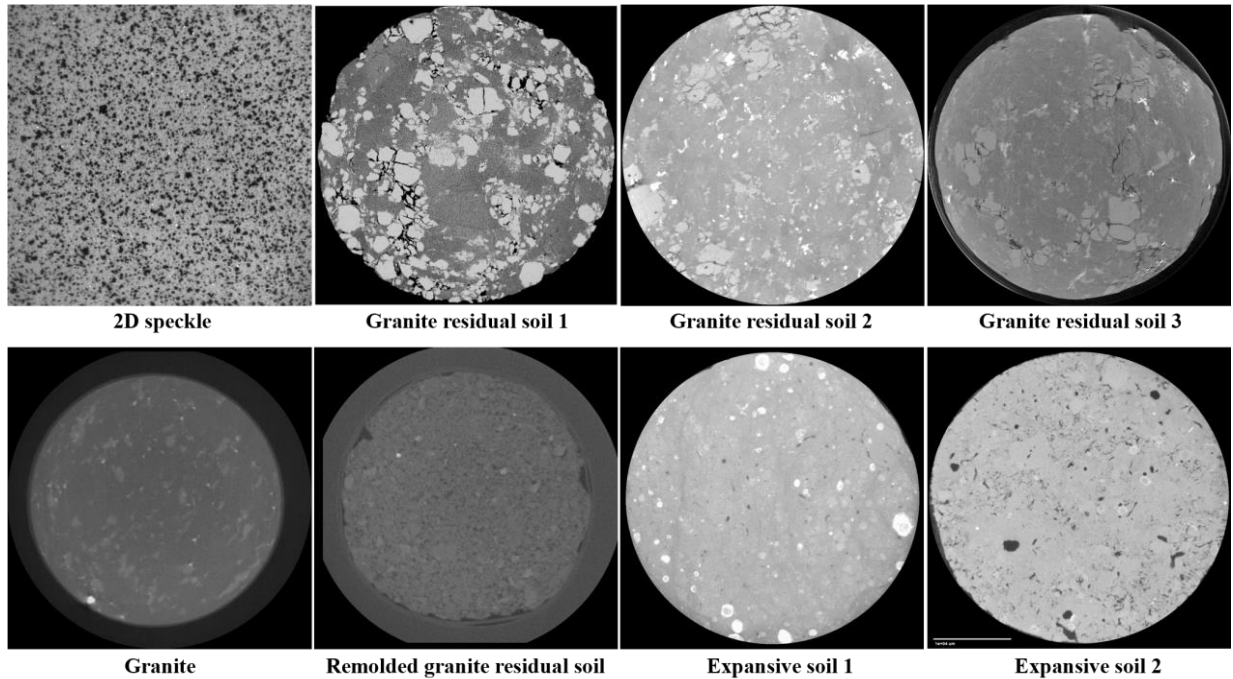

**Fig. 1.** Artificially 2D speckle and CT scan slices of different geomaterials

It is feasible to begin by calculating the displacement fields of adjacent correlated volume images, as depicted in Fig. 2, and then merge the multiple displacement fields to produce the final displacement field. Although this method can effectively mitigate errors resulting from decorrelation, it still may encounter the following two issues:

a) Even when computing the volume images of adjacent loading states, there may still be decorrelation problems due to substantial differences in deformation. Increasing the number of CT scan points can help combat this challenge. However, if the CT scan points are not dense enough, potentially due to the high cost of CT testing, significant deformation may persist between adjacent loading states.

b) If a substantial error occurs in the preceding calculation step and is not adequately corrected, these errors will compound and escalate during the displacement field merging process, resulting in significant inaccuracies in the final merged displacement field.

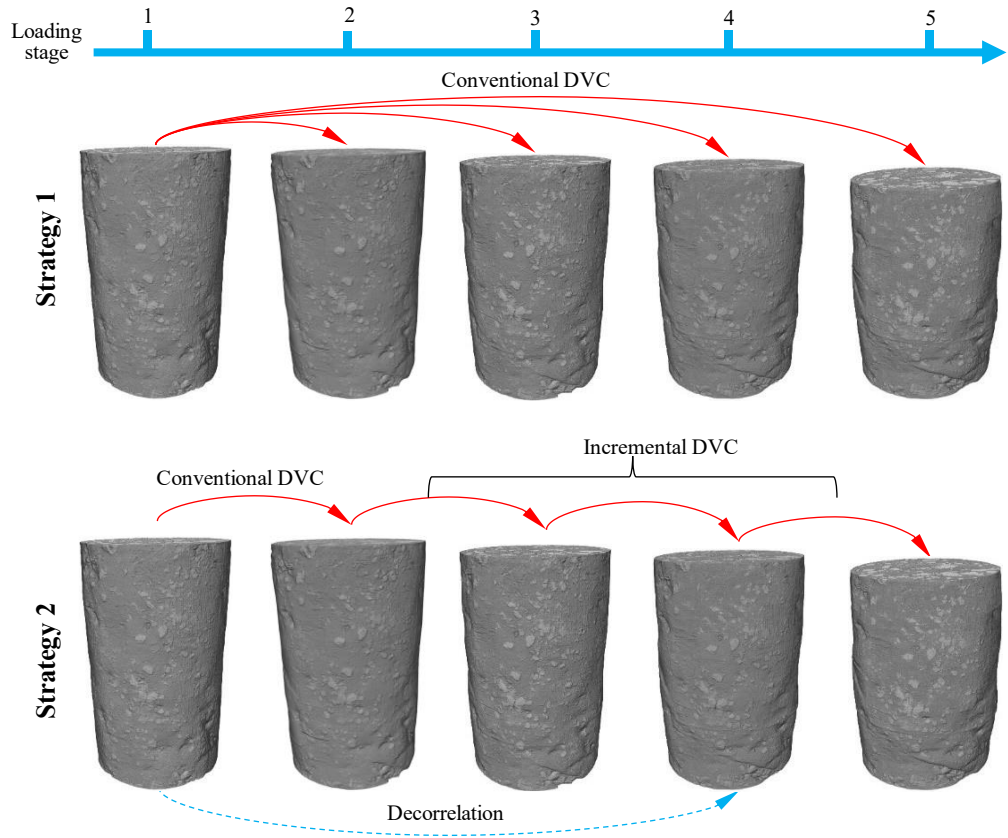

**Fig. 2.** Two typical DVC calculation strategies

Currently, it is challenging to obtain the three-dimensional displacement or strain fields inside a specimen non-destructively through existing experimental measurement methods (such as laser, strain gauges, optical fibers, etc.). This means that it is difficult to correct and validate the accuracy of the displacement fields calculated by DVC using these auxiliary methods. Additionally, as shown in Fig. 1, the grayscale values and imaging noise in the volumetric images reconstructed through CT scanning may not conform to some kinds of statistical models. Therefore, it is difficult to establish an effective displacement field correction method based on existing error theories. Although Rouwane et al. [23] implemented displacement field calculations for materials lacking texture using automatic finite element analysis combined with regularization techniques, it may be difficult to accurately segment key constituents (such as particles) in geotechnical materials (such as soil-rock mixture). Furthermore, after the deformation of the specimen, minerals in the material such as granular soils may exhibit behaviors like fragmentation, bonding, and rotation, posing challenges in structural modeling. Therefore, this method may not be suitable for these types of geotechnical materials.

However, it is worth noting that if the displacement field calculated by DVC is accurate enough, then transforming the volume image from the initial state to a certain loading state through the displacement field should result in a volume image that is similar to the deformed volume image. Specifically, the grayscale residual between them is small. Therefore, based on this principle, we propose a displacement field correction method based on image matching. This is an empirical method, and with the following conditions for its applicability:

a) Poor quality of CT speckles.

b) Complex and large deformations. In this case, the numerical values of the displacement field are large, and the subvoxel accuracy of the displacement field can be ignored.

c) The displacement field of the specimen roughly satisfies the deformation compatibility conditions (for example, the influence of microcracks generated after specimen deformation on the matching calculation of DVC is limited).

As shown in Fig. 3, the self-correction strategy for displacement fields based on images follows the specific calculation process outlined below:

a) Utilize the conventional DVC method to calculate the three-dimensional displacement field **D** between the two volume images (**G**, **F**) before and after deformation.

b) Preprocess the obtained three-dimensional displacement field from the previous step, and correct anomalies to ensure it satisfies the deformation compatibility conditions.

c) Perform spline interpolation on the above results three times to obtain a dense displacement field result.

d) Transform the initial state volume image **G** to a certain loading state **G′** based on the dense three-dimensional displacement field obtained in the previous step.。

e) Use the DVC method again to calculate the three-dimensional displacement field **ΔD** between **G′** and **F**.

f) Update the final displacement field，**D′** = **D** + **ΔD**.

In the calculation process, there is no need to add other calculation parameters or readjust the original DVC

calculation parameters. Furthermore, in the DVC calculation, the IC-GN algorithm is used to match subvolumes, employing a local DVC method.

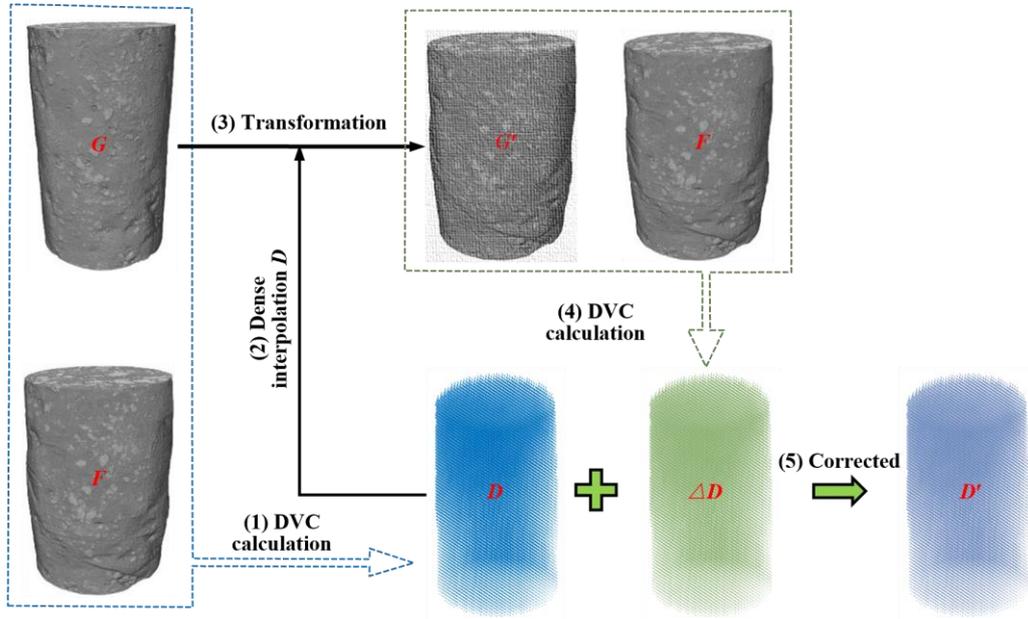

**Fig. 3.** Schematic diagram of the displacement field correction calculation process

**Evaluating the accuracy of displacement field**

(1) Estimation of background noise in reconstructed CT volume image

During the CT volumetric imaging process, the grayscale of each voxel in the post-deformation volume image will inevitably change. The calculation of residual grayscale fields needs to be considered and corrected for the influence of changes in grayscale in the target volume image [30]. However, for cases of complex and large deformations, it can be challenging to accurately obtain the displacement field in regions with complex and large deformations through image matching algorithms. As a result, analyzing grayscale residuals directly from the three-dimensional displacement field calculated through DVC may be limited. However, according to the principles of CT imaging, the grayscale in CT volumetric images is closely related to the density inside the material. The differences in grayscale values in CT imaging of areas with the same material composition can be considered as caused by imaging noise. Therefore, the imaging noise of CT volumetric images can be estimated based on the distribution characteristics of grayscale residuals for the same material properties.

The schematic diagram for estimating imaging noise in CT volumetric images is shown in Fig. 4, with analysis steps as follows:

a) Transform the reference volumetric image from the initial state to a certain deformed state by the three-dimensional displacement field calculated by DVC.

b) Using the results of material composition analysis (particle analysis tests) as a reference standard, perform multi-threshold segmentation on the CT volume images and label them to obtain a composition labeling matrix.

c) Take the difference between the CT volume images before and after deformation to obtain grayscale residuals. Then, overlap the threshold segmented images and label areas with the same component attributes.

d) Based on the results of the previous labeling step, calculate the mean grayscale residual of areas with the same composition and use this value as the estimated imaging noise value for the CT volumetric image.

Conventional DVC analysis struggles to perfectly match local speckles before and after deformation under complex and large deformation conditions. The above method only considers the mean grayscale residuals within the reasonable region of the statistical calculations. Therefore, this method can effectively mitigate the impact of displacement field calculation errors.

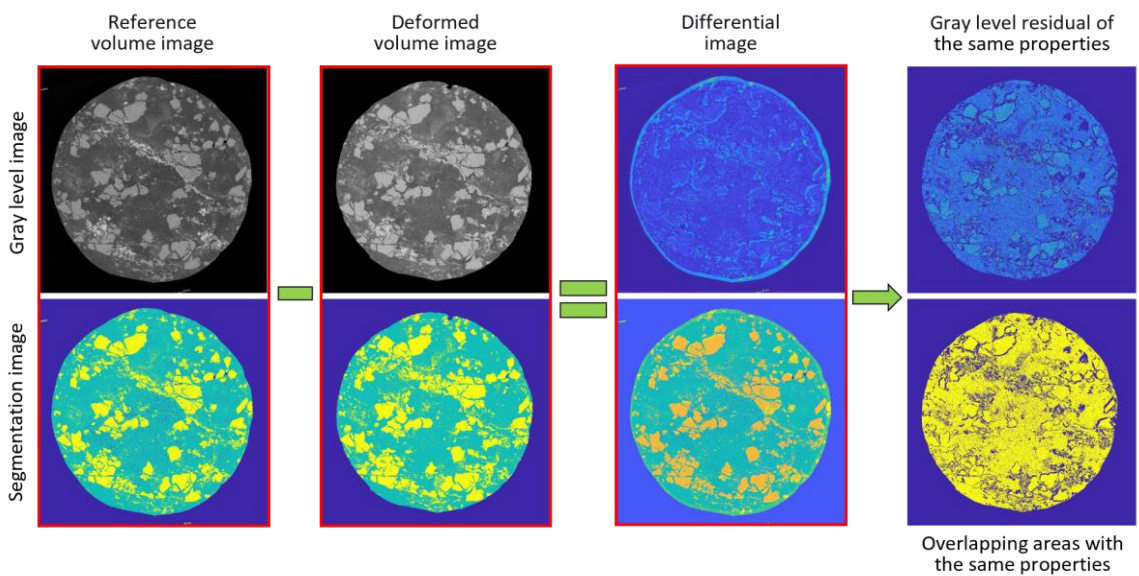

**Fig. 4.** Schematic diagram for background residual calculation in CT reconstruction

Gray residue correction based on background gray residue mean:

$$r' = r - \sum_{\{i,j\} \subset T} r_{0,ij} \Big/ M_T = r - \overline{r_0} \qquad (1)$$

where $r'$ is the corrected gray residue, $r$ is the directly calculated gray residue, $r_{0,ij}$ is the gray residue values in region $T$ with the same component composition, $M$ is the number of voxels in region $T$ with the same component composition, $\overline{r_0}$ is mean of the background gray residue.

(2) Estimation of the accuracy of the displacement field

If the displacement field calculated by DVC is accurate enough, then the grayscale residue should be very close to the background noise, or only in areas with significantly larger grayscale residue in locally damaged regions [31]. Conversely, if the accuracy of the displacement field calculated by DVC is not sufficient, then the grayscale residue will definitely be greater than the background noise. Therefore, the assumption is:

a) There are two kinds of materials with significant density differences in the material.

b) The initial DVC displacement field is close to the true value.

We try to use the following method to estimate the accuracy of the displacement field:

$$p = 1 - \frac{\overline{r'}}{|\overline{g_1} - \overline{g_2}|} \qquad (2)$$

where $p$ is the accuracy index of the displacement field, $\overline{g_1}$ the gray mean value represented as component one in volume image, $\overline{g_2}$ is the gray mean value represented as component two in volume image.

## Experimental results and discussion

### Triaxial CT test

We chose a natural granite residual soil shown in Fig.1 for a CT triaxial test to validate the proposed method's effectiveness. Fig. 5 illustrates that the granite residual soil contains numerous clay, quartz particles, and cracks, making its mesostructure complex and the distribution of each material component uneven. Based on the CT slices, reconstruction results, and grayscale statistics, the material's CT speckle quality is low. Prior to the CT triaxial test, a cylindrical sample of natural granite residual soil, with a diameter of 60 mm and a height of 120 mm, was prepared and positioned in the pressure chamber of the triaxial compression tester. The confining pressure was set at 200 kPa, and the drainage condition was established as undrained. CT scans of the sample at various strain levels were conducted to generate a sequence of volume images corresponding to axial strains of 4%, 6%, 9%, and 11.5%. The size of the volume images was 2000 × 2000 × 3000 voxels, as depicted in Fig. 2.

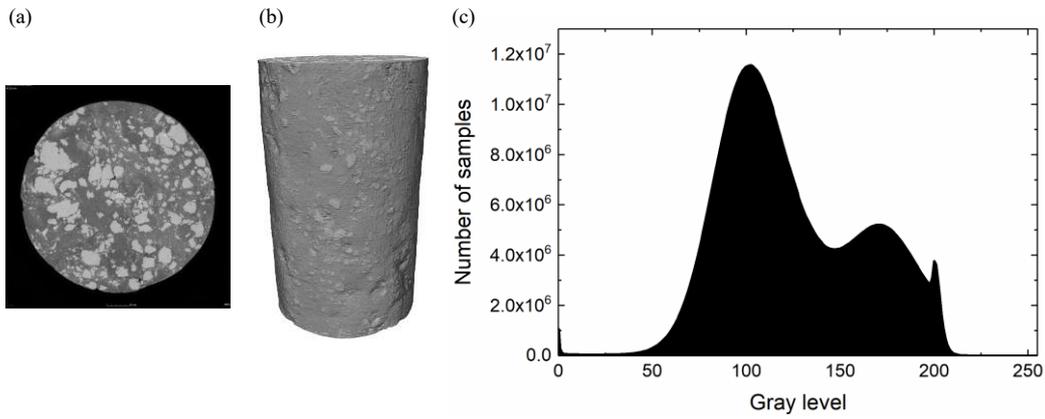

**Fig. 5.** The CT reconstruction results of the soil and the histogram of grayscale distribution

The DVC calculation parameters were set as follows: calculation step size of 10 voxels, subvolume size of 141 × 141 × 141 voxels, and subvolume matching using a first-order IC-GN algorithm. The displacement field calculation strategy employed the displacement superposition method shown in Fig. 2. All DVC analyses were performed using the in-house DVC software, known as inspire DVC (iDVC, https://github.com/lichengshengHK/iDVC).

As depicted in Fig. 6, we obtained the total displacement field and equivalent strain field (equation 3, octahedral strain) of the sample through DVC calculation. It is evident that when the axial strain of the sample reaches 9%, the maximum value of the total displacement field reaches 320 voxels, and the maximum equivalent strain value is 0.35. Upon reaching an axial strain of 11.5%, the maximum value of the total displacement field reaches approximately 630 voxels, with the maximum equivalent strain exceeding 0.4. This indicates that the sample undergoes significant deformation in the later stages of loading.

The equivalent strain formula is written as follows:

$$\bar{\varepsilon} = \sqrt{\frac{2}{9}[(\varepsilon_x - \varepsilon_y)^2 + (\varepsilon_y - \varepsilon_z)^2 + (\varepsilon_z - \varepsilon_x)^2 + 6(\varepsilon_{xy}^2 + \varepsilon_{yz}^2 + \varepsilon_{zx}^2)]} \qquad (3)$$

where the $\varepsilon_x$, $\varepsilon_y$, $\varepsilon_z$, $\varepsilon_{xy}$, $\varepsilon_{yz}$ and $\varepsilon_{zx}$ are the 6 components of strain.

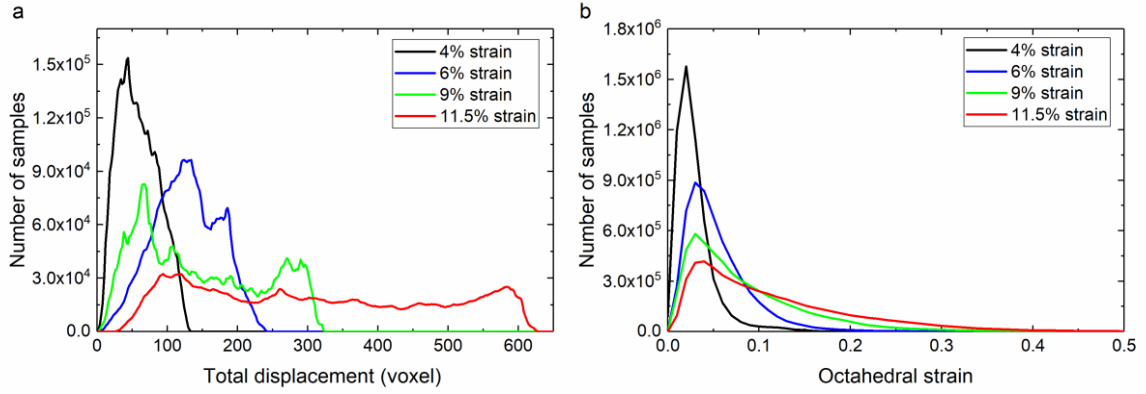

**Fig. 6.** The total displacement field and equivalent strain distribution curve at different loading stages

Furthermore, as illustrated in Fig. 7, strain localization phenomena gradually emerge within the sample as axial loading progresses. The strain values in these localized areas are notably higher than in other regions, showcasing an overall complex banded characteristic. Deformation in these localized areas is intricate, displaying significant shear deformation and potentially complex movements like the rotation of quartz particles. This complexity results in poor subvolume matching, ultimately impacting the accuracy of the displacement field. Therefore, in the presence of such extensive and complex deformations, the primary factor influencing the accuracy of the displacement field shifts from sub-voxel displacement to the integer voxels of the displacement field.

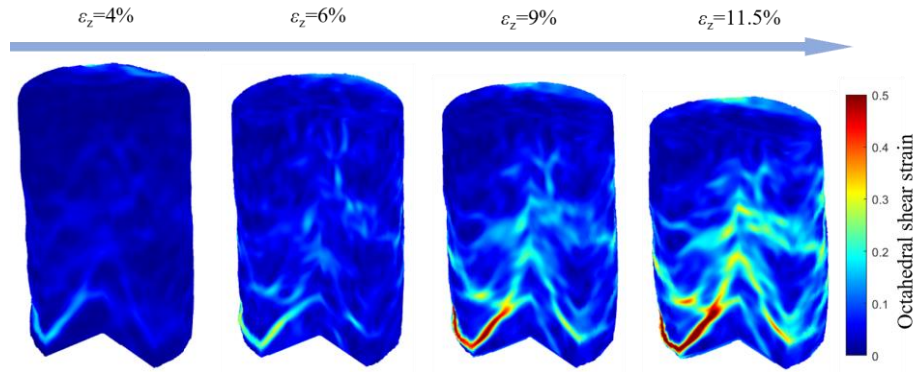

**Fig. 7.** The equivalent strain field of the uncorrected displacement field

**Gray residuals of uncorrected displacement field**

In most of the experimental measurements and analyses, whether by means of laser, strain gauge, or image measurement, we can not know the real displacement or deformation value of the specimen, especially for the three-dimensional displacement field and strain field inside the material. Therefore, in order to evaluate the accuracy of the three-dimensional displacement field calculated by the DVC method, the volume image of the specimen after deformation is used as a reference. Then the volume image in the initial state is transformed to a certain stage of loading according to the three-dimensional displacement field. Then the gray difference is made with the volume image after deformation, and the gray residual is used to evaluate the accuracy of the volume image matching. As shown in Fig. 8, we compare and analyze the DVC calculation results of 1-2, 1-3, 3-4, and 4-5, including gray residual and ZNCC.

Due to the intricate mesoscopic structure of granite residual soil, consisting of quartz particles of various sizes aggregated with clay, results in poor CT speckle quality within the sample. Additionally, the granite residual soil exhibits substantial plastic properties. As depicted in Fig. 7, complex strain localization phenomena manifest in the latter stages of loading. These factors contribute to the complexity of the displacement field. Even when employing DVC to compute the three-dimensional displacement field between two adjacent volume images, the accuracy of the resulting displacement field is subpar.

a) ZNCC cannot be used directly to evaluate the quality of matching results. When the axial strain of the sample is 4%, the overall ZNCC value is higher. However, there may be larger gray residuals in areas with higher ZNCC and smaller gray residuals in areas with lower ZNCC. These residuals do not follow a consistent distribution. The distribution disparity becomes more pronounced with the growing axial strain, leading to larger gray residuals in areas with higher ZNCC.

b) Referring to Fig. 7, in regions where sample deformation is minimal, a large ZNCC generally corresponds to a small gray residual. For instance, at an axial strain of 11.5%, this correlation is observed in the upper right corner region of the sample (Fig. 8, $x = 1000$ voxel). However, this pattern is not as noticeable in areas with significant sample deformation.

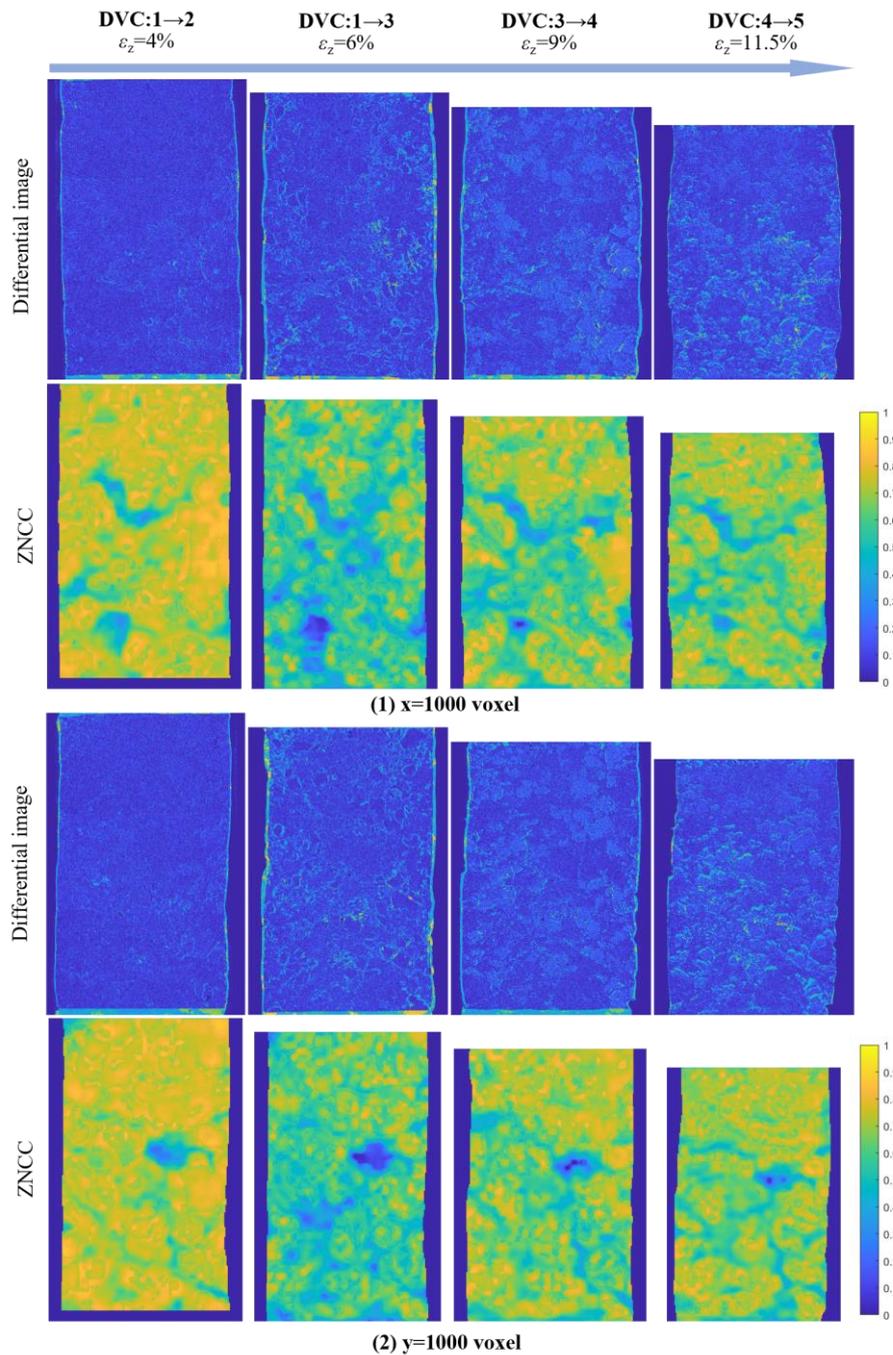

**Fig. 8.** Gray residual image and ZNCC image under incremental calculation strategy

Even when considering the three-dimensional displacement field of the specimen in two adjacent states, the gray residual analysis indicates a gradual decrease in displacement field accuracy with increasing axial strain. However, the behavior of ZNCC varies, with ZNCC values potentially increasing or decreasing.

**Gray residuals of corrected displacement field**

When stitching together the incremental displacement field results, significant errors in the accuracy of the displacement field at intermediate stages can directly impact the accuracy of subsequent displacement fields, potentially leading to noticeable errors. To address the challenges posed by calculating three-dimensional displacement fields with poor speckle quality, large deformation, and complexity, we have applied a proposed method to correct the three-dimensional displacement field. The following analysis is based on the results of the stitched displacement field. As depicted in Fig. 9 and 10, we have obtained grayscale residual slices and grayscale residual mean curves before and after correction. The findings indicate that the grayscale residuals have been effectively minimized.

The results indicate that:

a) At an axial strain of 4%, where the specimen's deformation is minimal and the displacement field accuracy is relatively high, the correction effect on the displacement field is not substantial. The mean grayscale residual value only decreases by approximately 3 to 4 gray values after the correction.

b) With an axial strain of 6%, there is a noticeable local increase in grayscale residual values in the displacement field before correction. After correcting the displacement field, the grayscale residual is significantly reduced, particularly in the range of equivalent strain between 0 to 0.15, indicating a marked improvement in displacement field accuracy.

c) At an axial strain of 9%, there is an overall reduction in the sample's grayscale residual after correction, especially in areas with lower equivalent strain. While some regions exhibit slight strain localization, the grayscale residual in these areas is also considerably decreased.

d) With an axial strain of 11.5%, significant strain localization is observed in the sample. Prior to the correction, the grayscale residual results indicated poor displacement field accuracy. However, post-correction, the grayscale residual in most areas is significantly reduced. Nevertheless, in localized strain regions, particularly in areas with equivalent strain exceeding 0.3, the correction effect on the displacement field is less effective.

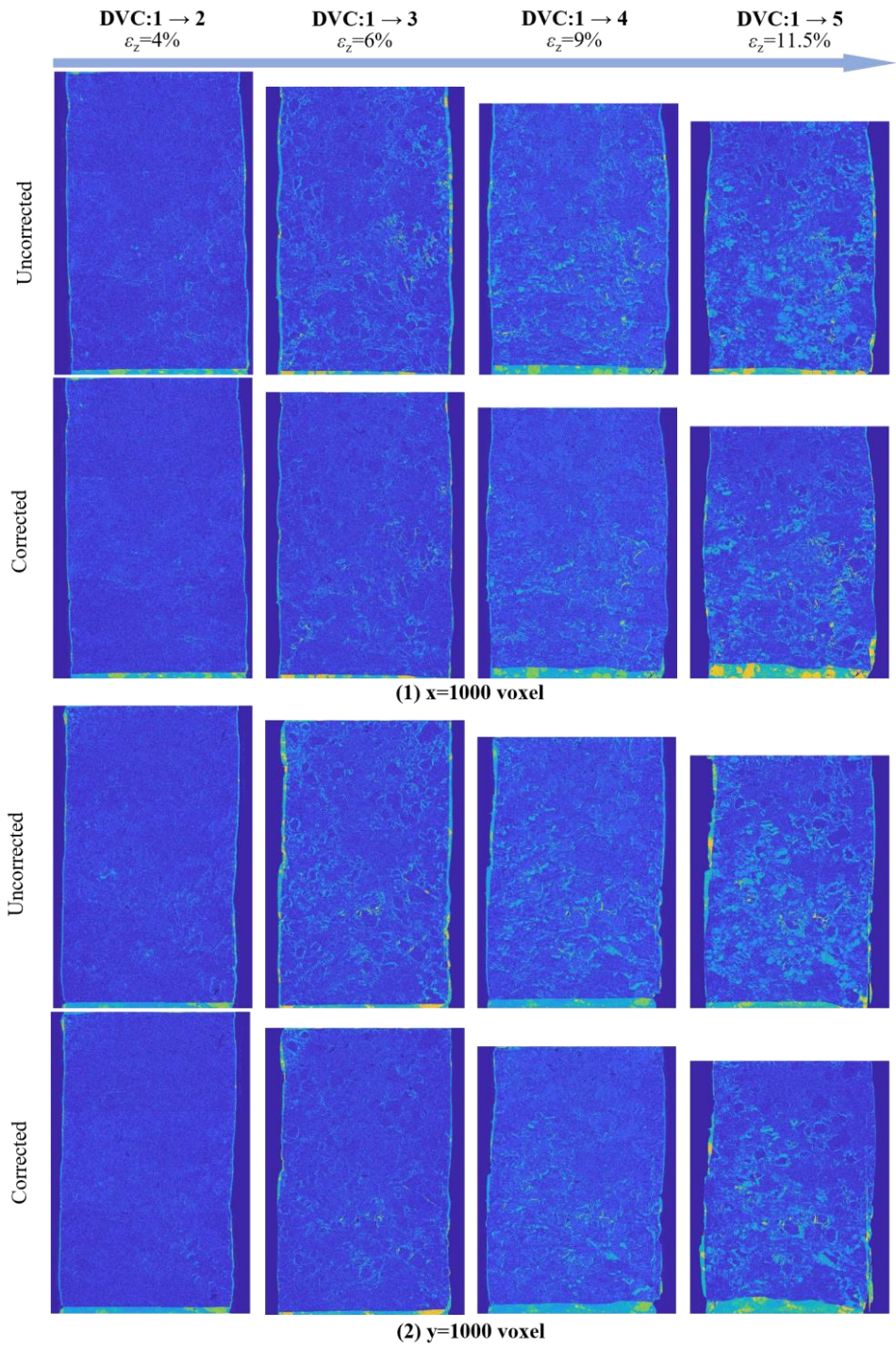

**Fig. 9.** Cracks of four samples of different loading stages

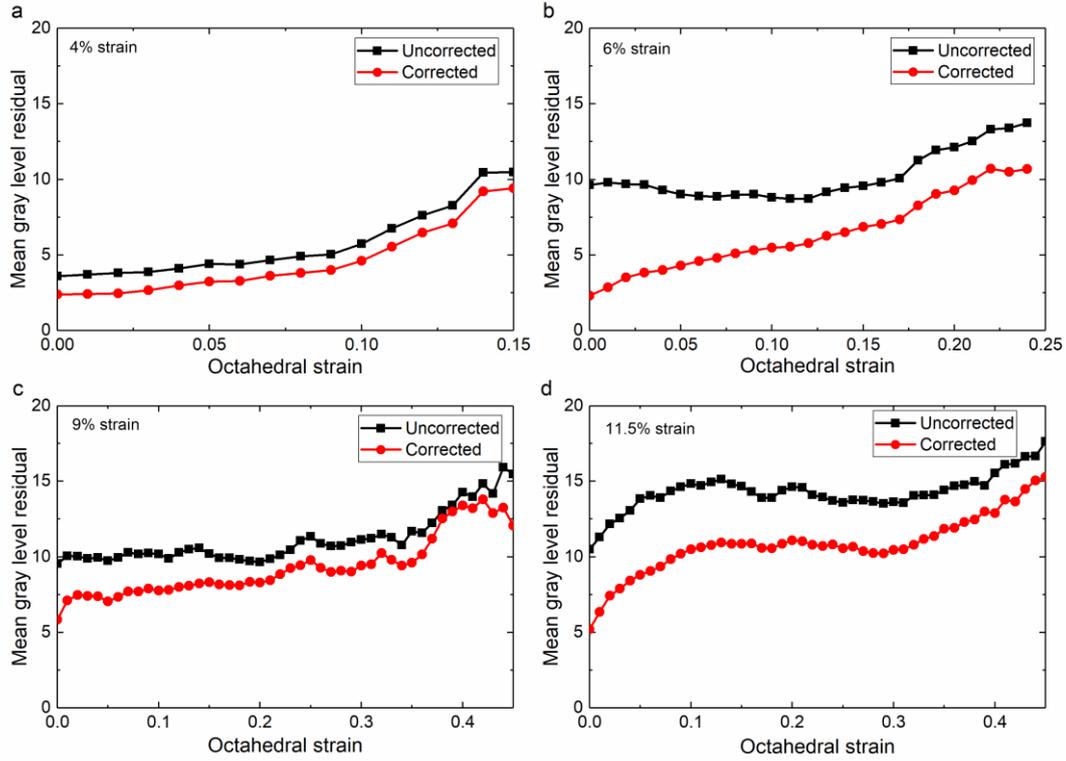

**Fig. 10.** Mean of grayscale residual under different strain conditions

**Discussion on error and accuracy**

In order to conduct a thorough assessment of the accuracy of the proposed method, we utilized the displacement field accuracy estimation technique that relies on grayscale residuals outlined in Section 2.2. Initially, we determined the segmentation threshold and mean grayscale values of both the soil and quartz particles within the sample using threshold segmentation. Subsequently, following the procedure depicted in Fig. 4, we calculated the mean residual value of the CT image background (Table 1). Finally, we determined the maximum accuracy limit of the displacement field based on equation 3.

According to Fig. 11, a 15% error margin (accuracy index of 0.85) is assumed for comparative analysis of accuracy indexes at different loading stages. At an axial strain of 4%, both the pre- and post-corrected displacement fields demonstrate high accuracy, with accuracy indexes exceeding 0.9 for equivalent strains below 0.1. Additionally, following the correction of the displacement field, there is an approximate 0.02 increase in accuracy.

Table 1 Statistical parameters of gray of volume image of granite residual soil

| NO | Axial strain (%) | Soil | | Quartz particles | | Mean gray residual of background |
|---|---|---|---|---|---|---|
| | | Gray threshold | Mean gray value | Gray threshold | Mean gray value | |
| 1 | 0 | 50-150 | 105.78 | 150-255 | 175.17 | — |
| 2 | 4 | 55-140 | 94.74 | 140-255 | 159.90 | 18.96 |
| 3 | 6 | 55-140 | 96.12 | 140-255 | 173.83 | 19.93 |
| 4 | 9 | 50-130 | 86.73 | 130-255 | 148.73 | 24.00 |
| 5 | 11.5 | 50-140 | 95.22 | 50-255 | 166.31 | 20.13 |

When the axial strain reaches 6%, although the overall accuracy index of the displacement field surpasses 0.85, significant enhancements in accuracy are observed post-correction, with accuracy indexes rising by approximately 0.05 to 0.1. Particularly in areas with lower equivalent strains, the improvement in the accuracy index is more noticeable.

At an axial strain of 9%, the accuracy index of the displacement field prior to correction remains around 0.85 for equivalent strains below 0.34. After applying the correction, the accuracy index improves by approximately 0.01 to 0.05. All accuracy indexes exceed 0.85 for equivalent strains below 0.34, indicating a significant enhancement in the precision of the displacement field.

At an axial strain of 11.5%, significant strain localization occurs within the specimen (as shown in Fig. 7), leading to accuracy indexes of the uncorrected displacement field dropping below 0.85. This suggests that the original displacement field is challenging to use for accurately analyzing the three-dimensional strain field. However, following the correction, the accuracy index of the displacement field within the range of equivalent strains less than 0.31 slightly exceeds 0.85, reaching an acceptable level.

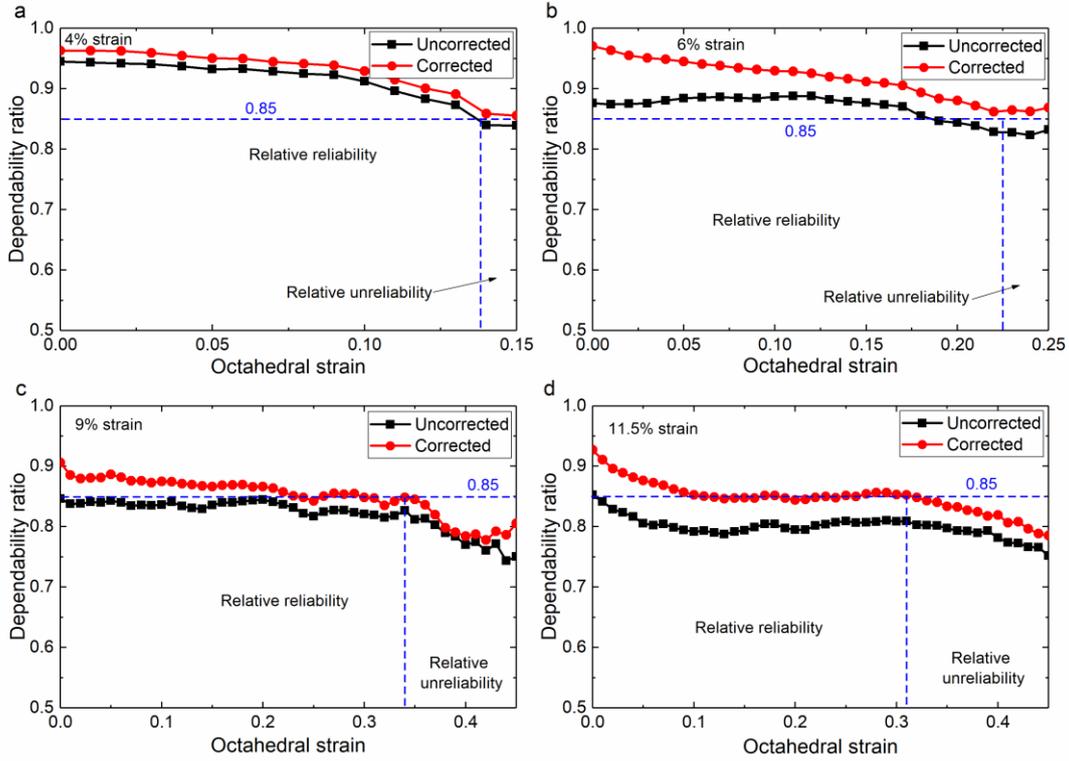

**Fig. 11.** The accuracy index before and after correction of displacement field

Furthermore, while the recommended displacement field correction method can effectively enhance the accuracy of the displacement field, its effectiveness in correction is constrained when the specimen's deformation is too significant. By combining information from Fig. 7 and Fig. 12, it is evident that at high axial strains, distinct strain localization phenomena such as shear bands manifest within the specimen. In areas of large deformation, certain quartz particles may experience significant rotations, fractures, or clustering, making it challenging to accurately calculate this type of deformation using the IC-GN algorithm. Consequently, the correction effects on the displacement field are limited. As depicted in Fig. 12, discrepancies in the displacement field within the strain localization regions of the specimen become more pronounced with increasing axial strain.

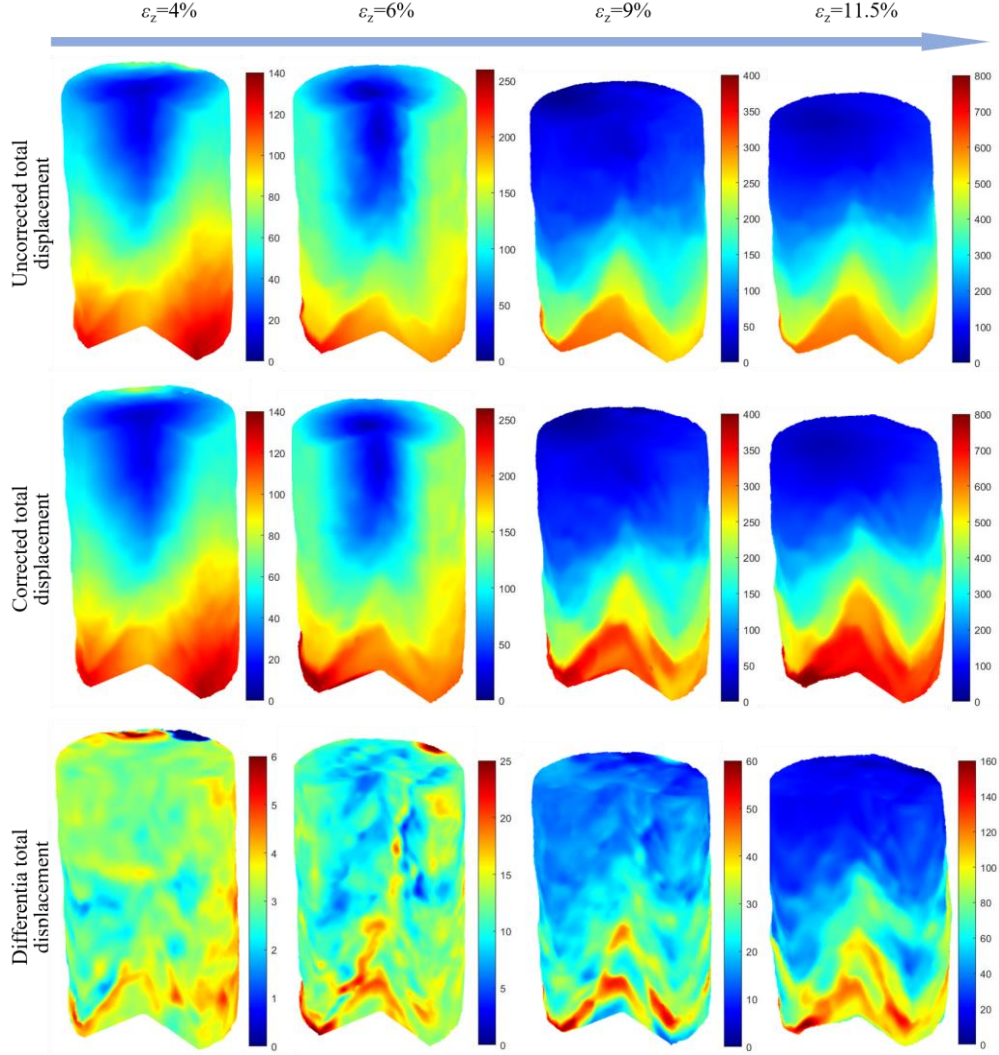

**Fig. 12.** Cloud map of the distribution of total displacement field before and after correction

# Conclusions

In this study, we did not establish a correction method for the DVC displacement field based on error theory. Instead, we analyzed the sources and characteristics of errors from the perspective of experimental measurements, combined with the speckle characteristics of CT volumetric images, and proposed a self-correcting strategy for the DVC displacement field based on image matching. This is an empirical method that does not introduce other parameters or require readjustment of DVC calculation parameters. Taking the CT triaxial test of granite residual soil as an example for validation, the results show that our proposed displacement field correction method can effectively improve the accuracy of the displacement field.   Additionally, based on the grayscale residuals of volumetric images, we established a simple method for estimating the accuracy of the displacement field, which can roughly evaluate the accuracy characteristics of the displacement field.

Moreover, the proposed method has some limitations, such as: 1) The method's corrective capacity is restricted when the specimen experiences substantial deformation, leading to poor accuracy of the DVC displacement field, especially in regions of strain localization. 2) Presently, the accuracy estimation method of the displacement field is appropriate for materials comprising two primary components and having an initial displacement field of acceptable accuracy. Further research is needed to devise methods that can be utilized for materials with multiple components.

## Declaration of competing interest

The authors declare that they have no known competing financial interests or personal relationships that could have appeared to influence the work reported in this paper.

## Acknowledgments

This work was supported by the Building Fund for the Academic Innovation Team of Shantou University (CN) (NTF21017), the Special Fund for Science and Technology of Guangdong Province in 2021 (STKJ2021181).